\documentclass[aps]{revtex4}
\begin{document}
\hfill{FERMILAB-FN-0298}

\title{\bf Very Big Accelerators as Energy Producers}

\author{R.R.~Wilson}
\affiliation{Fermi National Accelerator Laboratory\\Batavia, IL 60510 USA}
\date{August 9, 1976}

\maketitle

One consequence of the application of superconductivity to accelerator construction 
is that the power consumption of accelerators will become much smaller. This raises 
the old possibility of using high energy protons to make neutrons which are then 
absorbed by fertile uranium or thorium to make a fissionable material like plutonium 
that can be burned in a nuclear reactor~\footnote{Shades of E. O. Lawrence's MTA 
project! See {\it Atomic Shield}, page 425, by R. Hewlett and F. Duncan. I have also 
learned that there is a project in Canada to use an accelerator this way. See W. 
Metz, Science, July 23, 1976, Page 307.}.
The Energy Doubler/Saver
being constructed at Fermilab is to be a superconducting
accelerator that will produce 1000 GeV protons. The expected
intensity of about $10^{12}$ protons per second corresponds to a
beam power of about 0.2 MW. The total power requirements of
the Doubler will be about 20 MW of which the injector complex
will use approximately 13 MW, and the refrigeration of the
superconducting magnets will use about 7 MW. Thus the beam
power as projected is only a few orders of magnitude less
than the accele~ator power. But each 1000 GeV proton will
produce about 60,000 neutrons in each nuclear cascade shower
that is releaseq in a block of uranium, and then most of
these neutrons will be absorbed to produce 60,000 plutonium
a toms. Each of these when burned will Subsequently release
about 0.2 GeV of fission energy to make a total energy of
12,000 GeV (20 ergs) for each 1000 GeV proton. Inasmuch as
megawatts are involved, it appears to be worthwhile to
consider the cost of making the protons to see if there
could be an overall energy production.

A high energy proton accelerator seems at first glance
to be an unlikely device for producing energy in this manner
but high energy does have a few special advantages. One is
that once the protons have been injected and accepted by the
accelerator at the low energy of injection, then the subsequent
losses during the acceleration to high energy are essentially
zero. Another is that space charge limitations of intensity
become less important at high energy. A third advantage is
that the betatron oscillations of the beam are damped as the
energy increases. This means that the emittance of the beam
gets better as the energy grows and hence the beam can be
transferred to rings of smaller aperture (lower refrigeration
costs) as the energy and hence ring size increases. This
assumes that the accelerator consists of a series of magnet
rings of ever increasing radius. Alternatively, more protons
can be injected into a given aperture and accelerated. The
main point is that the number of neutrons produced is roughly
proportional to the beam power and this can be made large by
increasing both the intensity {\em and} the energy of the protons.
The intensity of an accelerator usually runs into a hard limit
imposed by space charge and resonance phenomenon, but the
energy can be increased without limit.

A disadvantage of very high proton energy is that the
relative number of neutrons produced in the nuclear shower
gets somewhat less as the energy of the initial proton
increases because the fraction of the electromagnetic component
increases with energy. This adverse effect could be vitiated
in part by accelerating deuterons (or particles of even
higher mass) so that the energy per nucleon is lower. In
any case, once a proton has gone through the expensive and
inefficient business of being produced and accelerated to
about 10 GeV energy, then all the energy possible should be
pumped into it during the efficient part of the acceleration
process that brings it to high energy.

Now let us consider the process a little more quantitatively.
We assume that the number of neutrons produced in a nuclear
shower in U$^{238}$ when $N$ protons of energy $E$ are incident is
proportional to the energy in the shower, i.e., to $aNE$,
where the constant $a$ will have a value of roughly 60 neutrons
per GeV, about half of which come from fission of U$^{238}$
Then the potential power $P$ produced overall when these neutrons are
nearly all absorbed to form plutonium is given by
\[P = 0.2 aNE - P_{0} - bNE,\]
where 0.2 is the energy per fission in GeV, $P_{0}$ is the power
required to run the accelerator when no protons are accelerated,
and $bNE$ is the power used by the RF system to accelerate the
protons. The constant $b$ is a measure of the inefficiency of
producing RF power from the electrical mains; it has a value
of roughly 2 although this might be improved to a value of
about 1.5.

For a given accelerator, if $0.2a$ is greater than $b$, as
it is for the Energy Doubler, then there will be a proton
intensity $N_{0}$ above which more power will be available from 
the plutonium burned than will be used by the accelerator.
This value is given by \[N_{0} \approx P_{0}/10E.\]
For the Energy Doubler at Fermilab plus all its injector
stages, $P_{0}$ will be roughly 20 MW, and $N_{0}$ then comes out to
be about $2 \times 10^{13}$ protons per second. This is about twenty
times the expected intensity - but it is far from being
unattainable. An intensity of $10^{13}$ protons per second will
make about 15 Megawatts of fission energy available; this
does not count the energy put into the accelerator. For an
overall production of 15 MW, an intensity of $3 \times 10^{13}$ protons
would be required; or more generally \[P \approx 10 (N-N_{0}) E.\]

Now the above calculations are pessimistic in that they
assume that the beam energy is thrown away. Furthermore
there will be considerable fission energy produced by fast
neutrons being absorbed in the U$^{238}$ during the course of the
cascade shower. Let us now assume that both of these forms
of energy are available for use. Mr. Andreas VanGinneken
has made a rough computer calculation following the shower
development and assuming that each U$^{238}$ fission leads to 3
Pu$^{239}$ nuclei, but that the Pu$^{239}$ fission is not used as a
source of neutrons. His results are tabulated in Table I
and illustrate the relative increase of the electromagnetic
component with energy. Remaking the calculations for the
rougher break-even intensity made above, we find $N_{0}$ to be
close to $10^{13}$ protons per second, an intensity which might
be attainable even in the Doubler by increasing the rate of
pulsing, or by resorting to a stacking process, or by adding
a second ring of superconducting magnets. He has also made
the same calculation at 100 and 300 GeV.

It appears then that an accelerator that is similar to
the Energy Doubler could be made to be energy productive.
Starting from the beginning, perhaps the injector system
could be made to use less ertergy and produce higher intensities.
One could also optimize the size of the rings so as to
reduce the refrigeration requirement and to raise the intensity
capability. An important improvement (and one assumed in
all the above) would be to make at least two magnet-rings
back-to-back so that the electric energy could oscillate from
one to the other rather than just being put back into the
power mains. This would double the intensity and would make
efficient use of the injection system, for one ring could be
loaded while the other was accelerating the protons previously
loaded into it. My guess is that the optimum energy of the
accelerator will be smaller, but this will depend on details
of the injection system and on the shape of the real estate
that is available. Values of $N_{0}$ for other energies are
given in Table I: they vary roughly inversely with the
energy but 100 GeV gives fractionally about 30\% more energy
per proton than does 1000 GeV.

Capital costs, of course, are equally significant. The
bare-bones accelerator might cost, as a very rough guess,
about \$200 million for a plant that would produce the fuel
to power a 100 MW fission plant. A larger installation
might of course cost relatively less per MW.

When the kind of intense proton beam considered above
is to be absorbed on a target, a serious problem arises.
Even with our present intensities at 400 BeV, i.e., about $10^{12}$
protons/sec, targets tend to disappear. One can imagine that
the target might be a slurry of uranium oxide and heavy
water. Each pulse of protons will last about twenty microseconds,
then the water will turn to stearn and start to
explode. This could move a piston in the classical manner
until the stearn has cooled (or has passed to a next stage)
and has been replaced again by the slurry. Such a dramatic
device need not be used; rather the stearn could be taken off
in the more usual manner of any power reactor and then used
for the generation of energy.

If undepleted uranium were to be used (or were slightly
enriched with Pu) then a magnification of the neutrons would
occur, but still without having a supercritical reactor.
This might bring the break-even intensity of protons needed
for power production down by another factor of ten, which
would then put it within range of the present Energy Doubler
accelerator at Fermilab.

There are probably better ways of producing plutonium,
but it does appear that it would be feasible to construct an
intense proton accelerator that would produce more energy
than it consumes. A further more careful study of the
accelerator as well as of the plutonium production in the
cascade showers would determine the optimum proton energy
for such a device. One happy result of all that intensity
would be that a truly magnificent neutrino source could be
produced!

I wish to thank Mr. A. Van Ginneken for making the nuclear
shower calculation.

\begin{center}
    \begin{tabular}{ | p{5cm} | l | l | l |}
    \hline
    \multicolumn{4}{|c|}{Table I} \\
    \multicolumn{4}{|c|}{Calculation of A. VanGinneken for protons on} \\
    \multicolumn{4}{|c|}{a large U$^{238}$ beam dump} \\
    \hline
           & 100 Gev & 300 Gev & 1000 Gev \\ \hline
     Ioniz. of Hadron Shower* & .34 & .29 & .23 \\ \hline
     E. M. Showers* & .32 & .40 & .48 \\ \hline
     U$^{238}$ Fission* & 2.2 & 1.9 & 1.6 \\ \hline
     Pu$^{239}$ Fission* & 13.8 & 11.8 & 10.2 \\ \hline
     Total* & 16.6 & 14.3 & 12.5 \\ \hline
     $N^{0}$ & $8.6\times10^{13}$ & $3.4\times10^{13}$ & $1.2\times10^{13}$ \\ \hline
     (Pu nuclei)/proton & 7000 & 18,000 & 62,000 \\ \hline
     Total Energy/1013 protons & 2.7 & 6.9 & 20 MW \\ \hline

    \end{tabular}
\end{center}

* Fraction of incident energy recovered: calculations assume
that each U238 fission makes three Pu239 nuclei. The
fission of Pu239 has not been considered as a source of neutrons.

\end{document}